\documentclass[11pt,twoside]{article}
\usepackage{atmp}
\copyrightnotice{2003}{7}{1}{23}

\begin{document}
\setlength{\parskip}{2ex} \setlength{\parindent}{0em}
\setlength{\baselineskip}{3ex}
\newcommand{\onefigure}[2]{\begin{figure}[htbp]
         \caption{\small #2\label{#1}(#1)}
         \end{figure}}
\newcommand{\onefigurenocap}[1]{\begin{figure}[h]
         \begin{center}\leavevmode\epsfbox{#1.eps}\end{center}
         \end{figure}}
\renewcommand{\onefigure}[2]{\begin{figure}[htbp]
         \begin{center}\leavevmode\epsfbox{#1.eps}\end{center}
         \caption{\small #2\label{#1}}
         \end{figure}}
\newcommand{\comment}[1]{}
\newcommand{\myref}[1]{(\ref{#1})}
\newcommand{\secref}[1]{sec.~\protect\ref{#1}}
\newcommand{\figref}[1]{Fig.~\protect\ref{#1}}
\newcommand{\mathbold}[1]{\mbox{\boldmath $\bf#1$}}
\newcommand{\mJ}{\mathbold{J}}
\newcommand{\momega}{\mathbold{\omega}}
\newcommand{\bz}{{\bf z}}
\def\bbbz{{\sf Z\!\!\!Z}}
\newcommand{\PP}{\mbox{I}\!\mbox{P}}
\newcommand{\bbbc}{\mbox{C}\!\!\!\mbox{I}}
\def\sl2z{SL(2,\bbbz)}
\newcommand{\bbbq}{I\!\!Q}
\newcommand{\be}{\begin{equation}}
\newcommand{\ee}{\end{equation}}
\newcommand{\bea}{\begin{eqnarray}}
\newcommand{\eea}{\end{eqnarray}}
\newcommand{\nn}{\nonumber}
\newcommand{\unit}{1\!\!1}
\newcommand{\half}{\frac{1}{2}}
\newcommand{\shalf}{\mbox{$\half$}}
\newcommand{\transform}[1]{
   \stackrel{#1}{-\hspace{-1.2ex}-\hspace{-1.2ex}\longrightarrow}}
\newcommand{\inter}[2]{\null^{\#}(#1\cdot#2)}
\newcommand{\lprod}[2]{\vec{#1}\cdot\vec{#2}}
\newcommand{\mult}[1]{{\cal N}(#1)}
\newcommand{\Bn}{{\cal B}_N}
\newcommand{\B}{{\cal B}}
\newcommand{\Beight}{{\cal B}_8}
\newcommand{\Bnine}{{\cal B}_9}
\newcommand{\Eman}{\widehat{\cal E}_N}
\newcommand{\C}{{\cal C}}
\newcommand{\Q}{Q\!\!\!Q}
\newcommand{\comp}{C\!\!\!C}

\noindent

\title{A note on E-strings}

\url{hep-th/0206064}

\author{Amer Iqbal}
\address{Theory Group, Department of Physics\\
University of Texas at Austin,\\
Austin, TX 78712, U.S.A.\\}
\addressemail{iqbal@physics.utexas.edu}

\begin{abstract}
We study BPS states in type IIA string compactification on a local
Calabi-Yau 3-fold which are related to the BPS states of the
E-string. Using Picard-Lefshetz transformations of the 3-cycles on
the mirror manifold we determine automorphisms of the K-theory of
the compact divisor of the Calabi-Yau which maps certain D-brane
configurations to a bound state of single D4-brane with multiple
D0-branes.  This map allows us to write down the generating
functions for the multiplicity of these BPS states.
\end{abstract}

\markboth{\it A NOTE ON E-STRINGS}{\it AMER IQBAL}

\newpage



\section{Introduction}

String dualities have played an important role in our
understanding of perturbative and non-perturbative phenomena in
supersymmetric theories. An interesting and important application
of dualities has been to the enumerative geometry of maps. BPS
amplitudes in string theory which are protected by supersymmetry
from quantum corrections have interesting mathematical
interpretations in the dual theory. Notable among these are the
${\cal N}=2$ topological string partition functions which count
curves of various genera \cite{AM,kodaira-spencer,KKV}, ${\cal
N}=1$ superpotential which counts maps from a disk with boundary
mapped to a Lagrangian cycles \cite{AV1,mayr,GJ, IP,JB,AAHV, DFG}
and the partition function of the ${\cal N}=4$ Vafa-Witten theory
which is the generating function of the Euler characteristic of
the instanton moduli spaces
\cite{vafa-witten,estring,ADE1,ADE2,ADE3,dijkgraaf1,dijkgraaf2,dijkgraaf3,
lozano1, lozano2, lozano3, lozano4,yoshioka3}.

In this paper we will discuss the partition functions of the
Vafa-Witten (VW) theory and its relation with the E-string
\cite{estring, SWC,mohri,HST} partition functions \cite{estring}. 
String dualities allow us to interpret the
partition function of the VW theory as the generating function for
the multiplicities of certain BPS states in the type IIA string
compactification on a Calabi-Yau threefold. These partition
functions have been evaluated for only a few cases such as $T^{4},
K3$ and the rational elliptic surface ${\cal B}_{9}$ ($\PP^{2}$
blown up at nine points)\footnote{In \cite{vafa-witten} a
conjecture for the rank two partition functions of the VW theory
on $\PP^{2}$ was also given. Also a conjecture about the $U(r)$
partition function ($r$ an odd prime) of VW theory on ${\cal
 B}_{9}$ was made in \cite{yoshioka3}}. In the case of ${\cal
B}_{9}$ the expressions obtained in \cite{estring}, partition
functions of E-string, were a sum, over the flux in the fiber
direction, of the partition functions $Z^{r}_{\Sigma}(\tau)$ of
the VW theory, \bea {\cal
Z}_{r,\Sigma}(\tau)=\frac{1}{r}\sum_{n=0}^{r-1}Z^{r}_{\Sigma+n\,F}(\tau)\,.
\eea Where $F$ is the class of the fiber and $\Sigma\in
H_{2}({\cal B}_{9},\bbbz)$ is such that $\Sigma\cdot F=0$.  The
purpose of this note is to show that Picard-Lefshetz
transformations of the 3-cycles in the mirror manifold $Y$ and the
map from $H_{*}(X)$ to $H_{3}(Y)$ allows us to map an arbitrary
D-brane configuration to a bound state of D2-brane and D0-branes. Therefore 
allowing us to compute $Z^{r}_{\Sigma}(\tau)$ without any
restriction on $\Sigma\cdot F$.

The configurations studied in \cite{estring} were bound states of
$Q_{4}$ D4-branes wrapped on ${\cal B}_{9}$ and $Q_{0}$ D0-branes.
Since ${\cal B}_{9}$ is an elliptically fibered manifold therefore
T-duality on the elliptic curve converts the D4-brane into a
D2-brane wrapped on the base of the fibration and converts the
D0-brane into D2-brane wrapped on the elliptic fiber. Thus the
D4-brane/D0-brane configuration is mapped to a D2-brane wrapping a
two cycle in ${\cal B}_{9}$.  Thus the problem of counting the
multiplicity of the BPS states given by D4-brane/D0-brane bound
states becomes the problem of counting certain curves in
Calabi-Yau threefold $X$ which can be done using mirror symmetry.
The configuration of D4-brane/D0-brane we started with, however,
is not the most general possible BPS configuration since one can
also have D2-brane wrapped on some curve. Such a configuration
with D2-brane wrapped on a curve, however, cannot be mapped to a
D2-brane wrapping some other curve using the T-duality on the
fiber of ${\cal B}_{9}$. We will show that since we can perform
$\sl2z$ transformations on the fiber we can map an arbitrary
configuration of D4/D2/D0-branes to a D2-brane wrapped on a curve
with some D0-branes.

The paper is organized as follows. In section two we review some facts
about the second homology lattice of ${\cal B}_{9}$, the structure of
the curve counting functions, and their relation with E-strings
partition functions and partition functions of VW theory. In section
three we discuss the action of $\sl2z$ and the generalized Weyl
transformations on the D-brane configurations. In section four we use
the transformations discussed in section three to write down partition
functions of VW theory. Some facts about the curves in ${\cal B}_{9}$
and the action of affine $E_{8}$ on the curves is discussed in
appendix A.

After this paper appeared we were informed by Kota Yoshioka that
theorem 6.15 and corollary 6.18 of his paper ``Twisted stability
and Fourier-Mukai Transform" (math.AG/0106118) may provide
mathematical evidence for the results of section three.

\section{Review of curve counting and E-string partition functions}
In this section we will review the relation between the curve counting
functions (genus zero topological string amplitude) and the E-string
partition functions \cite{estring}.

The Calabi-Yau space we will consider is total space of the canonical
line bundle over the rational elliptic surface. We will denote the CY
space by $X$ and the compact divisor by ${\cal B}_{9}$. ${\cal B}_{9}$
is an elliptically fibered surface with twelve degenerate fibers on
the base $\PP^{1}$. For this reason it is also known as
$\shalf$K3. For a more detailed construction of this CY from an
orbifold see \cite{estring}.
\subsection{$H_{2}({\cal B}_{9})$ and affine $E_{8}$}
The compact divisor ${\cal B}_{9}$ can be obtained from $\PP^{2}$ by
blown up nine points. We denote by $\{h,e_{1},\cdots,e_{9}\}$ a basis
of $H_{2}({\cal B}_{9},\bbbz)$ such that \bea h^{2}=1\,,\,\,h\cdot
e_{i}=0\,,\,\,e_{i}\cdot e_{j}=-\delta_{ij}\,. \label{interr} \eea
Here $h$ is the pullback of the generator of $\PP^{2}$ and $e_{i}$ are
the exceptional curves introduced by blowing up $\PP^{2}$ at nine
points. The canonical class is given by $3h-\sum_{i=1}^{9}e_{i}$ and
the homology classes orthogonal to it forms a codimension one lattice
isomorphic to the affine $E_{8}$ lattice. The real simple roots
$\{\alpha_{1},\cdots,\alpha_{8}\}$ and the imaginary root $F$ of this
affine $E_{8}$ are \bea
\alpha_{i}&:=&e_{i}-e_{i+1}\,,\,i=1,\cdots,7\,,\,\,\\ \nn
\alpha_{8}&:=&h-e_{1}-e_{2}-e_{3}\,,\\ \nn F &:=&
3h-\sum_{i=1}^{9}e_{i}\,.  \eea From eq(\ref{interr}) it follows that
the imaginary root and the simple roots defined above have the following
intersection numbers,\bea \alpha_{a}\cdot
\alpha_{b}=-(A_{E_{8}})_{ab}\,,\,\,\alpha_{a}\cdot
F=0\,,\,F^{2}=0\,,\,\,a=1,\cdots 8\,.\eea Where $A_{E_{8}}$ is the
$E_{8}$ Cartan matrix. It is convenient to define a different basis
$\{\omega^{1},\cdots,\omega^{8},B,F\}$ of $H_{2}({\cal B}_{9})$ such that,
\bea
\omega^{a}:=\sum_{b=1}^{8}(A^{-1}_{E_{8}})_{ab}\alpha_{b}\,,\,\,B:=e_{9}\,.
\eea The classes $B$ and $F$ are the homology classes of the base
$\PP^{1}$ and fiber $T^{2}$ respectively 
and satisfy, \bea B^{2}&=&-1\,,\,\,F^{2}=0\,,\,\,B\cdot F=1\,,\\ \nn
B\cdot \omega^{a}&=&F\cdot \omega^{a}=0\,,\,\,a=1,\cdots,8\,.  \eea In
this basis we can write any class $\Sigma\in H_{2}({\cal
B}_{9},\bbbz)$ as, \bea
\Sigma=\sum_{a=1}^{8}\lambda_{a}\omega^{a}+d\,B+s\,F\,,\mbox{where}\,\,
\lambda_{a}=\Sigma\cdot \alpha_{a}\,,\,d=\Sigma\cdot
F\,,\,\,s=\Sigma\cdot (B+F)\,.  \eea We will always use this basis to
represent homology classes and therefore will use the notation
$(\vec{\lambda},d,s)$ to represent the curve given above. For
$\Sigma_{i}=(\vec{\lambda}_{i},d_{i},s_{i})$ the intersection product
is given by \footnote{$\vec{\lambda}\cdot
\vec{\mu}=\sum_{a,b=1}^{8}\lambda_{a}(A^{-1}_{E_{8}})_{ab}\mu_{b}.$}
\bea
\Sigma_{1}\cdot\Sigma_{2} = - \vec{\lambda_{1}} \cdot
\vec{\lambda_{2}} - d_{1}\,d_{2} +d_{1}\, s_2 + d_{2} \, s_1 
\eea 
For an effective curve $\Sigma$ the virtual genus $g$ can be determined
from the Adjunction formula $\Sigma\cdot\Sigma = 2g -2 + d$.
The intersection of the curve with the canonical class, $d$, is called
the degree of the curve.

$E_{8}$ Weyl transformations map $H_{2}({\cal B}_{9})$ to itself such
that if $w\in W(E_{8})$ and $\Sigma=(\vec\lambda,d,s)$ then \bea
w(\Sigma)=(w(\vec{\lambda}),d,s)\,.  \eea Since the codimension one
lattice orthogonal to the canonical class is affine $E_{8}$ lattice we
also have affine $E_{8}$ Weyl transformations. These are
reflections in the roots $\widehat{\alpha}_{m}=\alpha+m\,F$ where
$\alpha$ is an $E_{8}$ root and $m\in \bbbz$. These affine Weyl transformation are such
that if $\Sigma_{1}=(\vec\lambda_{1},d_{1},s_{1})$ and
$\Sigma_{2}=(\vec\lambda_{2},d_{2},s_{2})$ are related by such a
transformation then \bea \vec\lambda_{2}\in
w(\vec\lambda_{1})+m\,d\Gamma_{E_{8}}\,,\,\,d_{1}=d_{2}=d\,,\,\,\Sigma_{1}^{2}=\Sigma_{2}^{2}\,.
\eea This implies that for a curve of fixed degree and genus we only
need to consider curves with the highest weight vector for affine
$E_{8}$ at level $d$.  Two curves $\Sigma_\lambda$ and $\Sigma_\mu$ of
degree $d$ and genus $g$, with $E_8$ weights $\lambda$ and $\mu$
related as $\mu=w(\lambda)+d\lambda'$ can be transformed into each
other by an affine $E_8$ Weyl-transformation.  To see this consider two curves
$\Sigma_{\lambda}=(\vec\lambda,d,l_{\lambda})$ and $\Sigma_{\mu}=(\vec\mu,d,l_{\mu})$ of genus $g$ and degree $d$ with
$E_{8}$ weight vectors $\lambda$ and $\mu$ respectively.
Since  $\Sigma_{\lambda}^{2}=\Sigma_{\mu}^{2}=2g-2+d$ therefore
\bea
l_{\mu} = l_{\lambda}+ {1\over 2d} (\mu^2 - \lambda^2) \eea It
follows from $\mu=w(\lambda)+d\lambda'$ that one can write
$\mu=w(\lambda+d\lambda'')$ for some
 $\lambda''=\sum
s^i\alpha_i$, where $s^i$ are integers and $\alpha_i$ are the
$E_8$ simple roots. Now perform the following affine $E_8$
Weyl-transformation:\footnote{ $(\alpha,0,n)=\alpha+nF$, where $\alpha$ is an $E_{8}$ root and $n\in \bbbz$.}
\begin{eqnarray}
w_{(\alpha_8,\bar{m}^8)}\circ
w_{(\alpha_8,m^8)}\circ\ldots\circ
w_{(\alpha_1,\bar{m}^1)}\circ w_{(\alpha_1,m^1)};\qquad
\hbox{with} \quad \bar{m}^i-m^i=s^i,
\end{eqnarray}
on the curve $\Sigma_\lambda = (\lambda,d,l_\lambda)$ to obtain:
\begin{eqnarray}
\Bigl(\lambda+d\lambda''\,,d\,,\, l_\lambda + {1\over 2d} \{
(\lambda + d \lambda '')^2 - \lambda^2 \} \Bigr) &=&
\Bigl(\lambda+d\lambda''\,,d\,,\, l_\lambda + {1\over 2d} \{ \mu^2
- \lambda^2 \} \Bigr)\cr &=& \Bigl(w^{-1} (\mu)\,,d\,,\, l_\mu
\Bigr) .
\end{eqnarray}
This differs from the curve corresponding to $\mu$ by just a  Weyl transformation of $E_8$. As a
consequence we find that it is sufficient to study curves with dominant 
$E_{8}$ weight vector because any other curve
can be obtained from these by an affine $E_8$ Weyl-transformation.
Also this implies that the
multiplicity of a curve depends only on its degree, genus and
the equivalence class of its $E_8$ weight vector.

\subsection{Counting curves}
 Consider the generating function for the multiplicities of degree $n$
curves.  \bea \widetilde{Z}_{n}(\omega)=\sum_{C\in H_{2}({\cal
B}_{9},\bbbz),C\cdot F=n} {\cal N}(C)e^{2\pi i (C\cdot \omega)}\,,
\eea where ${\cal N}(C)$ is the Euler characteristic of the moduli
space of class $C$ if $C$ represents a holomorphic curve and $\omega$
is the K\"ahler form given by \bea
\omega=\sum_{a=1}^{8}m_{a}\omega^{a}+\tau(B+F)+\phi F\,.  \eea so that
\bea m_{a}=\omega\cdot \alpha_{a}\,,\,\,\tau=\omega\cdot
F\,,\,\,\phi=\omega\cdot B\,.  \eea Since the degree of the curve,
which is also the level of the $E_{8}$ weight vector, is invariant
under the affine Weyl transformations therefore
$\lambda:=w(\lambda^{a})+n\,\Gamma_{E_{8}}$ $\forall \lambda\in
\Gamma_{E_{8}}$. Here $\{\lambda^{a}~|~a=1,\cdots F(n)\}$ is the set
of dominant weights at level $n$. We denote by $\Delta^{a}$ the set of
vectors in $\Gamma_{E_{8}}$ related to $\lambda^{a}$ by $E_{8}$ Weyl transformations and a shift in $n\Gamma_{E_{8}}$. Thus \bea
\widetilde{Z}_{n}(\omega)=\sum_{a=1}^{F(n)}\widetilde{Z}^{a}_{n}(\omega)\,,
\eea where \bea \widetilde{Z}^{a}_{n}(\omega)=\sum_{d_{C}=n,
\lambda_{C}\in \Delta^{a}} {\cal N}(C) e^{2\pi i (C\cdot
\omega)}\,. \eea Since ${\cal N}(C)={\cal N}(w(C))$, where $w\in
W(E_{8})$, therefore,\bea
\widetilde{Z}^{a}_{n}(\omega)=\sum_{d_{C}=n, \lambda_{C}\in
\Delta^{a}} {\cal N}(C) \sum_{w\in W(E_{8})}e^{2\pi i (w(C)\cdot
\omega)}\,. \eea The curves of degree $n$ are given by \bea
C=\sum_{a=1}^{8}\lambda_{a}\omega^{a}+nB+\{\frac{2g-2+n+n^{2}+\lambda^{2}}{2n}\}F\,,\,\,
\eea where $g=1+\frac{C^{2}-n}{2}$ and therefore, \bea
\widetilde{Z}^{a}_{n}(\omega)=e^{2\pi i n\phi}\sum_{g,\lambda\in
\Delta^{a}}{\cal N}_{n}^{a}(g) e^{2\pi i\tau
(\frac{2g-2+n+n^{2}}{2n})}\,\sum_{w\in W(E_{8})}e^{2\pi i
(\frac{\tau}{2n}\lambda^{2}-(w(\lambda),m))}\,. \eea In the above
expression ${\cal N}_{n}^{a}(g)$ is a function of degree $n$, genus
$g$ and the dominant weight $\lambda^{a}$ since all vectors in
$\Delta^{a}$ are Weyl equivalent to $\lambda^{a}$. Thus \bea
\widetilde{Z}^{a}_{n}(\omega)&=&q^{n/2}e^{2\pi i n\phi}\sum_{g}{\cal
N}_{n}^{a}(g)q^{\frac{2g-2+n}{2n}}\,\sum_{\lambda\in
\Delta^{a}}\sum_{w\in W(E_{8})}q^{\frac{\lambda^{2}}{2n}}\,e^{-2\pi i
(w(\lambda),m)}\,,\\ \nn &=& q^{n/2}e^{2\pi i n
\phi}\widehat{Z}^{a}_{n}(\tau)P_{n,a}(\vec{m},\tau)\,. \eea Where \bea
\widehat{Z}^{a}_{n}(\tau)&=&\sum_{g}{\cal
N}_{n}^{a}(g)q^{\frac{2g-2+n}{2n}}\,,\label{genusexpansion}\\\nn
P_{n,a}(\vec{m},\tau)&=&\sum_{\lambda\in \Delta^{a}}\sum_{w\in
W(E_{8})}q^{\frac{\lambda^{2}}{2n}}\,e^{-2\pi i (w(\lambda),m)}\,.
\eea
The functions $\widehat{Z}_{n}^{a}(\tau)$  can be determined from
the genus zero topological string amplitude, $F_{0}(\omega)$. $F_{0}(\omega)$
can be calculated  using local mirror symmetry as in \cite{estring}.

\subsection{E-string partition functions}
The E-string partition functions capture the BPS states of M5-branes
wrapped on $\shalf\mbox{K3}\times S^{1}$ with some momentum in the
$S^{1}$ direction\cite{estring, bonelli1, bonelli2, bonelli3}.  The theory on $\shalf$K3 is twisted ${\cal N}=4$
Super Yang-Mills. Of the three possible twists of the ${\cal N}=4$
Yang-Mills the one that arises here was studied in \cite{vafa-witten}.
It was shown in \cite{vafa-witten} that the partition function of this twisted theory is the generating function of the Euler characteristic of the instanton moduli spaces.

To relate the E-string partition function to the partition
function of the ${\cal N}=4$ twisted theory consider a bundle $V$
of rank $r$ with $\Sigma$ as the 2-cycle dual to the first Chern
class and instanton number $l$, \bea \mbox{ch}_{0}(V)=r,
\,\,c_{1}(V)=\Sigma\,,\,\,\mbox{ch}_{2}(V)=k:=\shalf\,\Sigma\cdot\Sigma-l\,.
\label{charge} \eea Let ${\cal M}(r,\Sigma,l)$ be the moduli space
of the bundle with above Chern classes and $e^{r}_{\Sigma}(l)$ be
the Euler characteristic of the moduli space. We define the
generating function for the Euler characteristic as
\cite{vafa-witten} \bea Z^{r}_{\Sigma}(\tau):=\sum_{l}
e^{r}_{\Sigma}(l) q^{-\langle
V,V\rangle/2r}=q^{-\frac{r}{2}}\sum_{l}e^{r}_{\Sigma}(l)q^{\Delta(l)}\,,\,\,\,\,q=e^{2\pi
i \tau}\,. \label{ins} \eea Where $\langle
V,V\rangle=r^{2}-\Sigma\cdot \Sigma+2rk$ and
$\Delta(l)=l-\frac{r-1}{2r}\Sigma \cdot \Sigma=-\frac{\langle
V,V\rangle}{2r}+\frac{r}{2}$.  Define, \bea {\cal
Z}_{r,\vec{\lambda}}(\tau):=\frac{1}{r}\sum_{s=0}^{r-1}Z^{r}_{(\vec{\lambda},0,s)}(\tau),
\eea then ${\cal Z}_{r,\vec{\lambda}}(\tau)$ are the E-string partition
functions. Note that the E-string partition function are given by
generating function of instanton moduli space with first Chern
class of degree zero. This is also the moduli-space of a
configuration of D4-branes, D2-branes and D0-branes where the
number of D4-branes is equal to the rank of the bundle, the
D2-brane is wrapped on the 2-cycle dual to the first Chern class
and the number of D0-branes is equal to the instanton number.
However, since ${\cal B}_{9}$ is an elliptically fibered surface
therefore T-duality on both fibers maps a wrapped D4-brane to a
D2-brane wrapped on the base $B$, a D0-brane becomes a D2-brane
wrapping the fiber $F$ and a D2-brane wrapped on a curve $nB+kF$
maps to $n$ D4-branes wrapped on ${\cal B}_{9}$ with $k$
D0-branes. Thus if the degree of the first Chern class (the cycle
on which the D2-brane is wrapped) is zero the entire configuration
of D4/D2/D0-branes can be mapped to a curve. Thus in this case the
twisted Yang-Mills partition function can be related to curve
counting function. In \cite{estring} this fact was used to relate
the E-string partition functions, which are given by generating
function of instanton moduli spaces with degree of the first Chern
class zero, to the genus zero topological string amplitude as
follow, \bea (-1)^{r-1}{\cal
Z}_{r,\lambda^{a}}(\tau)=\frac{q^{-n/2}}{n^{3}}\delta_{\lambda^{a},0}+\widehat{Z}^{a}_{r}(\tau)\,.
\label{rel} \eea The first term which is only present for
$\lambda^{a}=0$ is due to the fact that $nB$ is not a holomorphic
curve for $n>1$ (it does not satisfy the adjunction formula) and
therefore the corresponding term is not present in
$\widehat{Z}_{n}^{0}(\tau)$ but genus zero topological string
amplitude does contain such a term due to contributions coming
from multiple cover of $B$.

The transformation property of ${\cal Z}_{r,\lambda}(\tau)$ under
S-duality can be easily determined from that of
$Z^{r}_{\Sigma}(\tau)$ \cite{vafa-witten}, \bea
Z^{r}_{\Sigma}(-\mbox{$\frac{1}{\tau}$})=(-1)^{r}\,r^{-5}\tau^{-6}\sum_{\Sigma'\in
H_{2}({\cal B}_{9},\bbbz_{r})} e^{\frac{2\pi i}{r} \Sigma\cdot
\Sigma'}Z^{r}_{\Sigma'}(\tau)\,. \label{sduality} \eea This
implies that
\begin{eqnarray}
Z^{\,r}_{(\lambda\,,\,0\,,\,s)}(\mbox{$-\frac{1}{\tau}$})&=&(-1)^{r}\,r^{-5}\tau^{-6}
\sum_{{(\mu\,,\,d\,,\,s')\atop \in H_{2}({\cal B}_{9},\bbbz_{r})}}
e^{\frac{2\pi
i}{r}(-\lambda\cdot\mu+s\,d)}Z^{\,r}_{(\mu\,,\,d\,,\,s')}(\tau)\,,\\
\nn &=&(-1)^{r}\, r^{-5}\tau^{-6}\sum_{\mu\in{\Gamma_8/
r\Gamma_8}} \,\,\sum_{d,s'=0}^{r-1}\,e^{\frac{2\pi i}{r}
(-\lambda\cdot\mu+s\,d)}Z^{\,r}_{(\mu\,,\,d\,,\,s')}(\tau)\,,
\\ \nn
&=&(-1)^{r}\, r^{-5}\tau^{-6} \sum_{\mu\in{\Gamma_8/ r\Gamma_8}}
\,\,e^{-\frac{2\pi i}{r}\lambda\cdot\mu}\Bigl\{\sum_{s'=0}^{r-1}
Z^{\,r}_{(\mu\,,\,0\,,\,s')}(\tau)+\\ \nn
&&\sum_{d=1\atop s'=0}^{r-1}
\,\,e^{\frac{2\pi i}{r}s\,d}Z^{\,r}_{(\mu\,,\,d\,,\,s')}
(\tau)\Bigr\}\,,
\\ \nn &=& (-1)^{r}\,r^{-5}\tau^{-6}
\sum_{\mu\in{\Gamma_8/ r\Gamma_8}}
\,\,e^{-\frac{2\pi i}{r}\lambda\cdot\mu}\Bigl\{\,r\,{\cal Z}_{r,
\mu}(\tau)+\\ \nn
&&\sum_{d=1\atop s'=0}^{r-1}\,e^{\frac{2\pi
i}{r}s\,d}Z^{\,r}_{(\mu\,,\,d\,,\,s')}(\tau)\Bigr\}\,.
\label{ssduality}
\end{eqnarray}
Summing over $s$ on both sides of the equation and noting that the second term in the right hand side gives no
contribution we find
\bea
{\cal
Z}_{r\,,\,\lambda}(\mbox{$-\frac{1}{\tau}$})=(-1)^{r}\,r^{-4}\tau^{-6}
\sum_{\mu\in{\Gamma_8/ r\Gamma_8}} \,e^{-\frac{2\pi
i}{r}\lambda\cdot\mu}{\cal Z}_{r\,,\, \mu}(\tau). \eea

This implies that \bea r^{4}\tau^{6}{\cal
Z}_{r,\,0}(\mbox{$-\frac{1}{\tau}$})-{\cal Z}_{r,\,0}(\tau)=
\sum_{a=1}^{F(r)}C_{a}{\cal Z}_{r,\lambda^{a}}(\tau)\,.  \label{sumsum}\eea
Where $C_{a}$ is the number of elements in $\Delta^{a}$ and
$\sum_{a=1}^{F(r)}C_{a}=r^{8}$. Also in writing the left hand side of eq(\ref{sumsum})
we have used eq(\ref{rel}) and the fact that multiplicity of the
curve only depends on the degree, genus and the class of the
weight vector. In \cite{estring} partition functions were
calculated for $r=1,2,3,4$ by solving the holomorphic anomaly
equation.

{\bf Rank one and two:} Here we list the functions ${\cal
Z}_{r,\lambda^{a}}$ for $r=1$ and $r=2$ \cite{estring}, \bea {\cal
Z}_{1,0}(\tau)&=&\eta^{-12}(\tau)\,,\\ \nn {\cal
Z}_{2,0}(\tau)&=&-\frac{1}{24\eta^{24}}(E_{2}P_{0}(\tau)
+(\theta_{3}^{4}\theta_{4}^{4}-\frac{1}{8}\theta_{2}^{4})(\theta_{3}^{4}+\theta_{4}^{4}))\,,\\
\nn {\cal
Z}_{2,\lambda_{even}}&=&-\frac{1}{24\eta^{24}}(\frac{1}{135}E_{2}P_{even}(\tau)-\frac{1}{8}\theta_{3}^{4}(\theta_{3}^{4}+\theta_{4}^{4}))\,,\\
\nn {\cal
Z}_{2,\lambda_{odd}}&=&-\frac{1}{24\eta^{24}}(\frac{1}{120}E_{2}P_{odd}(\tau)-\frac{1}{8}\theta_{2}^{4}E_{4})\,.
\eea There are three dominant weights of $E_{8}$ at level 2,
$\lambda=0, \lambda_{even}$ and $\lambda_{odd}$. Where
$\lambda_{even}^{2}=4, \lambda^{2}_{odd}=2$ and $C_{even}=135,
C_{odd}=120$. Definition of various function appearing in the
equation above are given in appendix C.

For $r=2$ it is possible to write ${\cal Z}_{2,\lambda_{even}}$
and ${\cal Z}_{2,\lambda_{odd}}$ in terms of ${\cal
Z}_{2,0}(\tau)$. Define $F(\tau):=2^{4}\tau^{6}{\cal
Z}_{2,0}(\mbox{$-\frac{1}{\tau}$})-{\cal Z}_{2,0}(\tau)$ then from
eq(\ref{sumsum}) it follows that \bea\nn \sum_{a=1,2}C_{a}{\cal
Z}_{2,\lambda^{a}}(\tau)&=&F(\tau):= -\frac{1}{24\eta^{24}(\tau)}\{
\widehat{E}_{2}(\tau)(P_{0}(\frac{\tau}{4})-P_{0}(\tau))-\\ \nn
&&2^{4}(\theta_{3}^{4}\theta_{2}^{4}-\frac{1}{8}\theta_{4}^{8})(\theta_{3}^{4}+\theta_{2}^{4})
-(\theta_{3}^{4}\theta_{4}^{4}-\frac{1}{8}\theta_{2}^{8})(\theta_{3}^{4}+\theta_{4}^{4})\}\,.\nn
\eea Then from eq(\ref{genusexpansion}) it follows that ${\cal
Z}_{2,\lambda_{odd}}(\tau)$ has an expansion in integer plus half
powers of $q$ whereas ${\cal Z}_{2,\lambda_{even}}(\tau)$ has an
expansion in integer powers of $q$. Therefore, \bea\nn {\cal
Z}_{2,\lambda_{odd}}(\tau)&=&-\shalf(F(\tau)-F(\tau+1))\,,\\
&=&\frac{1}{48C_{odd}\eta^{24}}\{E_{2}(\tau)(E_{4}(\frac{\tau+1}{2})-E_{4}(\frac{\tau}{2}))+\\\nn
&&2^{4}(\theta_{4}^{4}\theta_{2}^{4}+\frac{1}{8}\theta_{3}^{8})(\theta_{4}^{4}-\theta_{2}^{4})
+2^{4}(\theta_{3}^{4}\theta_{2}^{4}-\frac{1}{8}\theta_{4}^{8})(\theta_{3}^{4}+\theta_{2}^{4})\}\,\\\nn
{\cal
Z}_{2,\lambda_{even}}(\tau)&=&-\shalf(F(\tau)+F(\tau+1))\,,\\\nn
&=&\frac{1}{48C_{even}\eta^{24}}\{E_{2}(\tau)(E_{4}(\frac{\tau+1}{2})+E_{4}(\frac{\tau}{2})-2E_{4}(\tau))+\\ \nn
&&2^{4}(\theta_{4}^{4}\theta_{2}^{4}+\frac{1}{8}\theta_{3}^{8})(\theta_{4}^{4}-\theta_{2}^{4})
-2^{4}
(\theta_{3}^{4}\theta_{2}^{4}-\frac{1}{8}\theta_{4}^{8})(\theta_{3}^{4}+\theta_{2}^{4})-\\ \nn
&&2(\theta_{3}^{4}\theta_{4}^{4}-\frac{1}{8}\theta_{2}^{8})(\theta_{3}^{4}+\theta_{4}^{4})\}\,.\nn
\eea Note that both ${\cal Z}_{2,even}$ and ${\cal Z}_{2,odd}$
have integer expansion whereas ${\cal Z}_{2,0}$ does not. The
non-integer expansion of ${\cal Z}_{2,0}$ is due to the fact that
it contains contribution from multiple covers with weight
$\frac{1}{n^{3}}$ whereas ${\cal Z}_{2,odd}$ and ${\cal
Z}_{2,even}$ do not have any such multiple cover contribution and
are sum over curves which are not multiples of some other curve.

\section{Generalized Weyl transformations}
We saw in the last section that the affine Weyl transformations
relating different classes in $H_{2}({\cal B}_{9})$ lead to an
important organizing principle. The partition functions are organized
in terms of level $n$ dominant weights with curves of same degree,
genus and class of the dominant weight having the same multiplicity.
In this section we show that there are other transformations which
act not only on the second homology but on $H_{0}\oplus H_{2}\oplus H_{4}$. We call these transformation generalized Weyl transformations since
they might be related to the Weyl transformations of the double loop algebra
$\widehat{E}_{9}$ studied in \cite{7-brane,weyl}.

These transformations can be understood as the mirror of the
Picard-Lefshetz transformations of the 3-cycles in the mirror
Calabi-Yau. Since ${\cal B}_{9}$ is elliptically fibered we can perform
$\sl2z$ transformations on the fiber. It was shown in \cite{HI} that
the $S$-transformation acts like the Fourier-Mukai transformation on the
Chern characters of a bundle and the $T$-transformation acts by tensoring
the bundle with a line bundle ${\cal O}(B)$.
Consider a bundle $V$ with Chern classes given by
\bea
\label{gbundle} \mbox{ch}_{0}(V)= r\,,\,\,\mbox{ch}_{1}(V)=
\Sigma=\{\vec\lambda,d,s\}\,,\,\,\,\mbox{ch}_{2}(V)=k\,. \eea Then
the $S$ and the $T$ transformation maps these charges to \cite{HI}
\bea S&:&\Bigl( r\,,\{ \lambda , d, s\} \,, k \Bigr)  \to \Bigl(
-d\,,\Bigl\{ w(\lambda )\, , r\,, -k + {\scriptstyle\half}\,
d\,\Bigr\} \,, s- {\scriptstyle\half} \,r \Bigr)\,, \label{S} \eea
\bea T&:&\Bigl( r\,,\{ \lambda , d, s\} \,, k \Bigr)  \to \Bigl( r
\,,\Bigl\{ w(\lambda )\, , d + r\,, s\,\Bigr\} \,, k + s- d -
{\scriptstyle\half} \,r \Bigr)\,. \label{T} \eea It is easy to
check that the product $\langle V,V\rangle$ is invariant under
both these transformations. The $\sl2z$ action on the rank $r$ and
the degree of the first Chern class $d$ is given by \bea
S&:&\,\,\begin{pmatrix} d\\ r \end{pmatrix}\mapsto\begin{pmatrix}r\\ -d\end{pmatrix}\,,\\ \nn
T&:&\,\,\begin{pmatrix}d\\ r\end{pmatrix}\mapsto \begin{pmatrix}d+r\\ r\end{pmatrix}\,. \eea The
$\sl2z$ transformation $-{\bf 1}$ changes the signs of all the
charges. Thus it follows that any bundle (D4/D2/D0-brane
configuration) can be mapped to a bundle with $r=0$ i.e., to a
curve.

The $E_{8}$ algebra in $H_{2}({\cal B}_{9})$ is enhanced by the
presence of $F$ (the imaginary root) to affine $E_{8}$ since
$F^{2}=0$. Instead of enhancing the algebra using $F$ we can also
use the charge vector $Q=(0,0,-1)$ (a 0-cycle) to enhance the
algebra since $\langle Q,Q\rangle=0$. Weyl transformations in
roots involving $Q$ give more non-trivial action on the charges
compared with the usual Weyl transformations as discussed in
Appendix C. \bea \label{fwr} G_{\pm}&:& \Bigl( 0\,,\{ \lambda , d
, s\} \,, k \Bigr) \mapsto \Bigl( 0\,,\{ \lambda , d, s\} \,, k
\pm  (\lambda, d)\Bigr)\,,\\ \nn w&:& \Bigl( 0\,,\{ \lambda , d,
s\} \,, k \Bigr) \mapsto \Bigl( 0\,,\{ w(\lambda) , d, s\} \,,
k\Bigr)\,,\,\,w\in W(E_{8})\,. \eea where $ (\lambda, d)  =d$ when
$\lambda=0$, and $(\lambda, d) = \mbox{gcd} (m, d)$ where $m$ is
the largest integer for which $\lambda \in m\Gamma_{E_{8}}$. Let
us consider an bundle $V_{1}$ such that degree of the first Chern
class of $V_{1}$ is zero. Then using (\ref{S}), (\ref{fwr}),
(\ref{S}), and the ${\bf -1}$ transformation successively we find
\begin{eqnarray}
\label{bundleweyl}
\mbox{ch}(V_1) =  \Bigl( r\,,\{ \vec\lambda , 0,
s\} \,, k \Bigr) &\mapsto &
  \Bigl( 0\,,\{ w(\vec\lambda) \,,\, r\,,\,
l+\frac{\lambda^{2}}{2}\} \,, s-\frac{r}{2} \,\Bigr)
\nonumber\\ &\mapsto & \Bigl( 0\,,\{ \vec\lambda\, , r,
l+\frac{\lambda^{2}}{2}\} \,, s-\frac{r}{2} +  (r,\vec\lambda) \Bigr) \\
&\mapsto & \Bigl( r\,,\{\vec\lambda\, , 0,
s+(r,\vec\lambda)\} \,, k\Bigr)\,.\nonumber
\end{eqnarray}
This implies that $e^r_{(\vec\lambda, 0, s)}(l)
= e^r_{(\vec\lambda, 0, s+ (r,\vec\lambda))}(l)$ and therefore
\bea
\label{feq}
Z_{( \vec\lambda\, , 0\,, \,s)}^{\,r} (\tau) = Z_{( \vec\lambda \,, 0\,,
\, s +(r,\vec\lambda))}^{\,r} (\tau) \,.
\eea
For the rank two case it follows that
\bea
Z^{2}_{(\lambda_{even},0,0)}(\tau)&=&Z^{2}_{(\lambda_{even},0,1)}(\tau)=
{\cal Z}_{2,\lambda_{even}}(\tau)\,,\\ \nn
Z^{2}_{(\lambda_{odd},0,0)}(\tau)&=&Z^{2}_{(\lambda_{odd},0,1)}(\tau)={\cal Z}_{2,\lambda_{odd}}
(\tau)\,.
\eea
This result (rank two case) was also given in \cite{yoshioka} where
${\cal Z}_{2,0}(\tau), {\cal Z}_{2,\lambda_{even}}(\tau)$ and ${\cal Z}_{2,\lambda_{odd}}(\tau)$ were also obtained by a rigorous mathematical analysis.

\section{Partition functions}
In this section we use the transformation discussed in the previous section
to obtain generating functions of the Euler characteristic of the instanton moduli spaces  for various ranks.

{\bf rank one:} This case has been discussed in several places
\cite{vafa-witten, estring} but we will consider it to illustrate the
method. Consider a rank one bundle such that\bea
\mbox{ch}_{0}(V)=1\,,\,\,\mbox{ch}_{1}(V):=\Sigma=\{\vec{\lambda},d,s\}\,,\,\,\mbox{ch}_{2}(V)=k=\shalf \Sigma\cdot \Sigma -l\,.
\eea Since zero is the only dominant vector at level one, by an
affine $E_{8}$ Weyl transformation we can set the $\vec\lambda$ in the equation above to zero. Also since the degree of the first Chern class and the rank is relatively prime therefore by an $\sl2z$ transformation we can map the above bundle to a bundle with rank zero. Thus the above bundle is mapped to $V_{1}$ such that \bea
\mbox{ch}_{0}(V_{1})=0\,,\,\,\mbox{ch}_{1}(V_{1})=\{0,1,l\}\,,\,\,\mbox{ch}_{2}(V_{1})=s-\shalf\,.
\eea Since the transformation $G_{-}$ allows us to change $s$ in the equation above by $-1$ therefore by repeated application of $G_{-}$ we get
\bea
\mbox{ch}(V_{2})=( 0,\{0,1,l\},-\shalf )\,.
\eea
Now the $S$ transformation maps it back to a rank one bundle $V_{3}$ with
\bea
\mbox{ch}(V_{3})=(1,\{0,0,0\},-l)\,.
\eea
And therefore $e_\Sigma^1 (l) = e^1_{(0,0,0)} (l)$.

To determine $e^{1}_{(0,0,0)}$ note that the bundle with these charge
can be realized a bound states of a D4-brane and $l$ D0-branes.  Thus
the moduli space is give by the $l-th$ symmetric product of
${\cal B}_{9}$ \cite{vafa-witten,estring}.\footnote{This is also the moduli
space of degree one curve $B+l\,F$ if we also include the flat
connections on this genus $l$ curve \cite{gopakumar}}. And therefore
the Euler characteristic is given by \bea
\sum_{l=0}^{\infty}\chi(Sym^{l}({\cal B}_{9}))q^{l}=\prod_{n=1}^{\infty}(1-q^{n})^{-12}=q^{1/2}\,\eta^{-12}(\tau)=\sum_{l=0}^{\infty}d(l)q^{l}\,.
\eea Thus $e^{1}_{(0,0,0)}(l)=d(l)$ and eq(\ref{ins}) implies that
\begin{eqnarray}
Z^{1}_{\Sigma}(\tau) &=& q^{-1/2}\sum_{l}
e^1_\Sigma (l)\, q^{l} \\\nn
&=&  q^{-1/2}\,\,q^{1/2}\eta^{-12}(\tau)=\eta^{-12}(\tau)\,.
\end{eqnarray}

{\bf rank r , degree $d$ with $\mbox{gcd}(d,r)=1$:} Now consider a
bundle $V$ with rank $r$ and degree of the first Chern class $d$ such that
$r$ and $d$ are relatively prime. Then since the $\sl2z$
transformation acts on $r$ and $d$ we can map this bundle to a bundle
with rank one. As discussed in the previous section, any bundle of rank one can be mapped to a bundle of rank 1 with zero first Chern class therefore,
\bea
\mbox{ch}(V)=(r,\Sigma=\{\lambda,d,s\},k)\mapsto \mbox{ch}(V_{1})=(1,\{0,0,0\},l_{1})\,.
\eea
Where $l_{1}$ can be determined by the fact that all transformation we have used preserve the product $\langle V,V\rangle =\langle V_{1},V_{1}\rangle $. This
gives ($k=\shalf \Sigma\cdot \Sigma -l$)
\bea
l_{1}=- \mbox{${\langle V, V\rangle \over 2}$} + \shalf= rl - \shalf\,(r-1)\Sigma^2 - \shalf (r^2-1)\,.
\eea
We then have ($d=F\cdot \Sigma$ and is relatively prime to $r$),
\bea
Z^r_{\Sigma} (\tau) &=& \sum_{l}e^r_{\Sigma} (l)
\, q^{-\frac{\langle V, V\rangle}{2r}} \\ \nn
&=& q^{-\frac{1}{2r}} \sum_{l_{1}=0}^{\infty} d(l_{1})
\, q^{\frac{l_{1}}{r}}\,.
\eea
Note that writing
$l_{1} = rl + \mu$ we see that the partition function for fixed rank
only depends on  $\mu$ (mod r) and  only $\Sigma^2$ enters in the
determination
of $\mu$.
Using the identity
\bea
\sum_{k=0}^{r-1}\omega_{r}^{(2g-2s)k} = r\delta_{g\equiv
s\,(\mbox{mod}\,\,r)}\,,\,\,\,\,\omega_{r}=e^{\frac{i\pi}{r}}\,\,.
\eea
We get
\begin{eqnarray}
Z^r_{\Sigma} (\tau)
&=&q^{-\frac{1}{2r}+\frac{\mu}{r}}\sum_{rl+\mu\geq 0}d(rl+\mu)\,q^{l}\\
\nn
&=&\frac{1}{r}\sum_{m=0}^{r-1}\omega_{r}^{(1-2\mu)m}\eta^{-12}(\mbox{$\frac{\tau}{r}$}+\mbox{$\frac{m}{r}$}).
\end{eqnarray}
Since $\mu=-\frac{1}{2}(r^{2}-1)-\frac{r-1}{2}\Sigma^{2}$ therefore
\bea
Z^{\,r}_{\Sigma}(\tau)=\frac{1}{r}\sum_{m=0}^{r-1}(-1)^{rm}\,e^{i\pi
m(r-1)\Sigma^{2}/r}\,\eta^{-12}(\mbox{$\frac{\tau}{r}$}+\mbox{$\frac{m}{r}$})\,.
\label{gcd}
\eea

{\bf rank r, degree $d=0$:} Consider a bundle of rank $r$ with degree
of the first Chern class equal to zero. These were the configuration considered
in \cite{estring}. Using the $S$-transformation we can map this bundle to a curve of degree $r$.

From eq(\ref{ssduality}) it follows that
\bea
\label{degzero} Z^{\,r}_{(\lambda\,,\,0\,,\,s)}(\tau)={\cal
Z}_{r\,,\,\lambda}(\tau)+(-1)^{r}\,r^{-5}\,\tau^{6}
\sum_{\mu\in{\Gamma_8/ r\Gamma_8}}\, \sum_{d=1\atop
s'=0}^{r-1}\,e^{\frac{2\pi i}{r}(-\lambda\cdot \mu+s\,d)}
Z^{\,r}_{(\mu\,,\,d\,,\,s')} (\mbox{$-\frac{1}{\tau}$}) \,. \eea
Specializing to prime rank $r=p$ and using eq(\ref{gcd}) one finds
\bea \label{fprr} Z^{\,p}_{(\lambda\,,\,0\,,\,s)}(\tau)={\cal
Z}_{p\,,\,\lambda}(\tau)-\frac{(-1)^{p}}{p^{11}} \,\Bigl(
\sum_{\mu\in{\Gamma_8/ r\Gamma_8}}\hskip-6pt e^{-2\pi
i\frac{\lambda\cdot \mu}{p}}\,\Bigr) \, \omega(p,s)\,
\eta^{-12}(p\tau) \eea where
\bea
\omega(p,s) = \sum_{d=1}^{p-1}\, e^{2i\pi \,sd/p} =
\begin{cases}
p-1, &\text{if $ s=0$} ; \\
 -1,  &\text{if  $s \not= 0$\,.}
\end{cases}
\eea
The sum in parenthesis in (\ref{fprr}) can be evaluated explicitly to obtain,
\bea
\label{fpr} Z^{\,p}_{(\lambda\,,\,0\,,\,s)}(\tau)={\cal
Z}_{p\,,\,\lambda}(\tau)-\frac{(-1)^{p}\,\delta_{\vec\lambda, p}}{
p^3} \, \, \omega(p,s)\, \eta^{-12}(p\tau)\,. \eea Here
$\delta_{\vec\lambda, p}$ vanishes unless $\lambda \in
p\Gamma_{E_{8}}$ in which case it takes the value one.

{\bf examples:} For the rank two case this give \bea
Z^{\,2}_{\Sigma}(\tau)&=&\shalf\{\eta^{-12}(\mbox{$\frac{\tau}{2}$})
+i^{\Sigma^{2}}\eta^{-12}(\mbox{$\frac{\tau}{2}+\frac{1}{2}$})\}\,,
\,\,\,\,\,\,\mbox{d}_{\Sigma}\in 2\bbbz+1\,,\\ \nn
Z^{2}_{(\vec\lambda\,,\,0\,,\,s)}(\tau)&=&{\cal
Z}_{2\,,\,\vec{\lambda}}(\tau)+(-1)^{s+1}\frac{\delta_{\vec\lambda,
2}}{8} \eta^{-12}(2\tau). \eea This result for the rank two case
was also obtained in \cite{yoshioka}. A conjecture about the
$U(r)$ partition function for $r$ an odd prime was made in
\cite{yoshioka3}. For rank three we have ($\mbox{d}_{\Sigma}=\Sigma\cdot F \neq 0\,(\mbox{mod}\,3)$)\bea \nn
Z^{\,3}_{\Sigma}(\tau)&=&\mbox{$\frac{1}{3}$}\{\eta^{-12}(\mbox{$\frac{\tau}{3}$})
-
e^{\frac{2i\pi}{3}\Sigma^{2}}\eta^{-12}(\mbox{$\frac{\tau}{3}+\frac{1}{3}$})
+e^{\frac{4i\pi}{3}\Sigma^{2}}\eta^{-12}(\mbox{$\frac{\tau}{3}+\frac{2}{3}$})\}\,,
\,\,\,\,\,,\\ \nn
Z^{\,3}_{(\lambda\,,\,0\,,\,s)}(\tau)&=&{\cal
Z}_{3\,,\,\lambda}(\tau)+{\delta_{\vec\lambda, 3}\over 27} \, \,
\omega(3,s)\, \eta^{-12}(3\tau)\,, \eea where $\omega(3,0) = 2$
and $\omega (3,1) = \omega (3,2) = -1$.

\subsection{A recursive scheme}
For prime rank $p$ the complete explicit solution can be obtained in
terms of $\eta^{-12}(\tau)$ and the E-string partition functions as shown
in the previous subsection.
For non prime rank $r$ there are three cases. When
gcd$(r,\mbox{d})=1$ we use (\ref{gcd}).  When the
gcd$(r,\mbox{d})=f>1$, we can map such bundles to rank $f$ (which is less than $r$) bundles. So these can be written in terms of lower rank
partition functions.  When the degree $d$ is zero we simply
use (\ref{degzero}) which expresses the partition function in
terms of an E-string partition function and (by the above
argument) partition functions of lower rank.

Consider a rank $r$ bundle $V$ of first Chern class $\Sigma$ such that
$\mbox{gcd}(r,\,\mbox{d}_{\Sigma})=f$. Then using (\ref{S}) and (\ref{T}) we can
map $V$ to a rank $f$ bundle
\bea
\Bigl( r , \{ \lambda , \mbox{d}_{\Sigma}, s\}, k\Bigr) \mapsto \Bigl( f , \{
\lambda ,0, s(l)\}, -\frac{\lambda^{2}}{2}-l'\Bigr)\,,
\eea
where
\bea
l'= r_0l + \mu\,, \quad r_{0}=\frac{r}{f}\,, \quad
\mu = -\frac{r-1}{2f}\Sigma^{2}-\frac{r^{2}}{2f}+
\frac{f}{2}-\frac{f-1}{2f}\lambda^{2}.
\eea
which implies that $e^{r}_{\Sigma}(l)=e^{f}_{(\lambda,0,s(l))}(l')$.
\begin{eqnarray}
Z^{\,r}_{\Sigma}(\tau)&=&
q^{-\frac{r}{2}-\frac{r-1}{2r}\Sigma^{2}}\sum_{l}e^{r}_{\Sigma}(l)q^{l}\\
&=&
q^{-\frac{r}{2}-\frac{r-1}{2r}\Sigma^{2}}\sum_{l}e^{r}_{(\lambda,0,s(l)}(r_{0}l+\mu)q^{l}
\\ \nn
&=&q^{\frac{\mu}{r_{0}}}\,q^{-\frac{f}{2r_{0}}-
\frac{f-1}{2fr_{0}}\lambda^{2}}\sum_{l}e^{f}_{(\lambda,0,s(l))}(r_{0}l+\mu)q^{l}
\end{eqnarray}
Let $h=(\vec\lambda,f)$ and define
\bea
s'(l)\equiv s(l) (\mbox{mod}\, h)
\eea
Since $s(l) = al+ b$ with integers $a,b$, the argument of  $s'$ can be restricted
to the integers mod $h$. Then we get
\begin{eqnarray}
Z^{\,r}_{\Sigma}(\tau)&=&
q^{\frac{\mu}{r_{0}}}\,q^{-\frac{f}{2r_{0}}-
\frac{f-1}{2fr_{0}}\lambda^{2}}\sum_{n=0}^{h-1}\,\,\,
\sum_{l=
n \,(mod\, h)} e^{f}_{(\lambda,0,s'(n))}(r_{0}l+\mu)q^{l}\\ \nn
&=&q^{\frac{\mu}{r_{0}}}\,
q^{-\frac{f}{2r_{0}}-\frac{f-1}{2fr_{0}}\lambda^{2}}\sum_{n=0}^{h-1}\sum_{l}
e^{f}_{(\lambda,0,s'(n))}(r_{0}hl+r_{0}n+\mu)q^{hl+n}\,.
\end{eqnarray}
It is not difficult to show that
\bea
q^{\frac{\mu}{r_{0}}-\frac{f}{2r_{0}}-\frac{f-1}{2fr_{0}}\lambda^{2}}\sum_{l}
e^{f}_{(\lambda,0,s'(n))}(r_{0}hl+r_{0}n+\mu)q^{hl+n}=\\ \nn
\frac{1}{r_{0}h}\sum_{m=0}^{r_{0}h-1}e^{\frac{2\pi i
m}{h}(\frac{r-1}{2r}\Sigma^{2}+\frac{r}{2}-n)}
Z^{f}_{(\lambda,0,s'(n))}(\frac{\tau}{r_{0}}+\frac{m}{r_{0}h})\,.
\eea
Thus we get
\bea
\label{recur}
Z^{r}_{\Sigma}(\tau)=\frac{1}{r_{0}h}\sum_{n=0}^{h-1}\sum_{m=0}^{r_{0}h-1}e^{\frac{\pi
i m r}{h}} e^{\frac{i \pi m }{h}(\frac{r-1}{r})\Sigma^{2}}e^{-\frac{2 \pi i m
n}{h}}Z^{f}_{(\lambda,0,s'(n))}(\frac{\tau}{r_{0}}+\frac{m}{r_{0}h})\,.
\eea

{\bf example:} Consider the case of rank $r=4$,
For $\mbox{d}_{\Sigma}\equiv
1\,,\,3 \,(\mbox{mod}\,4)$ it follows from eq(\ref{gcd}),
\bea\nn
Z^{\,4}_{\Sigma}(\tau)=\frac{1}{4}\{\eta^{-12}(\frac{\tau}{4})+
e^{\frac{3i \pi}{4}\Sigma^{2} }\eta^{-12}(\frac{\tau}{4}+\frac{1}{4})
+e^{\frac{6i \pi}{4}\Sigma^{2} }\eta^{-12}(\frac{\tau}{4}+\frac{2}{4})
 +e^{\frac{9i \pi}{4}\Sigma^{2} }\eta^{-12}(\frac{\tau}{4}+\frac{3}{4}) \}\,,
\eea
for
$\mbox{d}_{\Sigma}\equiv\, 2\, (\mbox{mod}\,4)$ it follows from eq(\ref{recur}),
\bea\nn
Z^{\,4}_{\Sigma}(\tau) = \frac{1}{2}\{{\cal Z}_{2\,, \lambda}(\frac{\tau}{2}) +
e^{\frac{3 i \pi}{4}\Sigma^{2}}{\cal Z}_{2\,,\lambda}(\frac{\tau}{2}+\frac{1}{2})\}
-\frac{\delta_{\vec\lambda,2}}{16}\{e^{\frac{3 i
\pi}{4}\Sigma^{2}}-e^{\frac{9 i
\pi}{8}\Sigma^{2}}\}\eta^{-12}(\tau+\frac{1}{2})\,,
\eea
for $\mbox{d}_{\Sigma}=0$ we get using eq(\ref{degzero})
\begin{eqnarray}
Z^{4}_{(\lambda\,,\,0\,,\,s)}(\tau)={\cal Z}_{\,4\,,\,\lambda}(\tau)+
\delta_{\vec\lambda ,2}\frac{(-1)^{s}}{2^{3}}{\cal Z}_{2\,,\,
\lambda/2}(2\tau)
-\delta_{\vec{\lambda},4}\,i^{s}\frac{(1+(-1)^{s})}{4^{3}}\eta^{-12}(4\tau)\,.
\end{eqnarray}

\section*{Acknowledgment}

\noindent I would like to thank Jacques Distler, Tam\'as Hauer and
Cumrun Vafa  for many valuable discussions. I would specially like
to thank Barton Zwiebach for many valuable discussions and
collaboration in the early part of the project. This research is
supported by the National Science Foundation under Grant No.
0071512.

\section*{Appendix A}

Let $C$ be a class of degree $n$ satisfying the adjunction formula
with genus $g$ then
\begin{eqnarray}
C&=&\sum_{i=1}^{8}\mu_{i}\omega^i+nB+ \label{Ccurve}
\frac{1}{2n}\{2g-2+n+n^{2}+\vec{\mu}^{2}\}F\,.
\end{eqnarray}
Every degree one class can be represented by a holomorphic curve.
This follows from the fact that $B+kF$ is holomorphic for $k \geq
0$ and this class is Weyl equivalent to
$\sum_{i=1}^{8}\lambda_{i}\omega^{i}+B+sF$ for any $\lambda\in
\Gamma_{E_{8}}$. The necessary and sufficient condition for a
class to be represented by a holomorphic curve is that its
intersection with exceptional curves be positive. Let $E$ be an
exceptional curve with $E_{8}$ weight vector $\lambda$, then
\begin{eqnarray}
C\cdot E
&=&\frac{n}{2}(\vec\lambda-\frac{\vec\mu}{n})^{2}+
\frac{2g-2+n-n^{2}}{2n}\,.
\end{eqnarray}
$C$ has a holomorphic representative if this intersection number
is nonnegative for all $\lambda\in \Gamma_{E_{8}}$.
\begin{eqnarray}
C\cdot E \geq  \frac{n}{2}\mbox{Min}_{\lambda\in \Gamma_{E_{8}}}(\lambda-\frac{\mu}{n})^{2}+
\frac{2g-2+n-n^{2}}{2n}\,.
\end{eqnarray}
Define
\bea
f_{n}(\mu)=\mbox{Min}_{\lambda\in \Gamma_{E_{8}}}(\lambda-\frac{\mu}{n})^{2}\,.
\eea
Note that if $\mu=w(\mu')+n\nu$ then $f_{n}(\mu)=f_{n}(\mu')$ where $\nu\in \Gamma_{E_{8}}$ and $w\in W(E_{8})$. This implies that $f_{n}(\mu)=f_{n}(\mu^{a})$, where $\mu$ belongs to the class $\mu^{a}$. \footnote{
Dominant weights $\lambda$ at level $n$ are given by
$\sum_{i=1}^{8}a_{i}\lambda_{i}\leq n\,$, where $a_{i}$ are
the Dynkin labels of $E_{8}$ and $\lambda_{i}\geq 0$. If $F(n)$ is the number of dominant weights at level $n$ then generating function is given by $\sum_{n=1}^{\infty}F(n)q^{n}=(1-q)^{-1}\,\prod_{i=1}^{8}(1-q^{a_{i}})^{-1}$.}

It follows that $g=g_{c}(n,\mu^{a})+n\,m$, $m\in \bbbz$ and \bea
g_{c}(n,\mu^{a}):=1+\shalf(n^{2}-n^{2}\,f_{n}(\mu^{a})-n)\,,
\label{genuseq}\eea
where $g_{c}(n,\mu^{a})$ is the genus of the degree $n$ curve with
weight vector $\mu^{a}$. Thus we see that $C\cdot E$ is
non-negative iff \bea g\geq g_{c}(n,\mu^{a})\,. \label{genuscon}
\eea If $C_{1}$ and $C_{2}$ are two curves satisfying the
adjunction formula and eq(\ref{genuscon}) then $C_{1}\cdot
C_{2}\geq 0$.  To see this note that since $C_{1}$ and $C_{2}$
satisfy the adjunction formula therefore \bea
C_{1}&=&\sum_{i=1}^{8}\mu_{i}\omega_{i}+n_{1}B+\frac{2g_{1}-2+n_{1}+n_{1}^{2}+\mu^{2}}{2n_{1}}F,~\\
\nn
C_{2}&=&\sum_{i=1}^{8}\nu_{i}\omega_{i}+n_{2}B+\frac{2g_{2}-2+n_{2}+n_{2}^{2}+
\nu^{2}}{2n_{2}}F. \eea The mutual intersection number is \bea
C_{1}\cdot C_{2}&=&-\lprod{\mu}{\nu}-n_{1}n_{2}+\frac{n_{1}}{2n_{2}}
\{2g_{2}-2+n_{2}+n_{2}^{2}+\nu^{2}\}+\\ \nn
&&\frac{n_{2}}{2n_{1}}\{2g_{1}-2+n_{1}+n_{1}^{2}+\mu^{2}\}.
\eea Since $C_{1}$ and $C_{2}$ obey eq(\ref{genuscon}) therefore, \bea
g_{1}\geq g_{c}(n_{1},\lambda^{a})\,\,\,\mbox{and}\,\,\,g_{2}\geq
g_{c}(n_{2},\lambda^{b}) \eea where $\lambda^{a}$ and
$\lambda^{b}$ are the level $n_{1}$ and level $n_{2}$ dominant
weights corresponding to the weight vectors $\mu$ and $\nu$
respectively. From these two conditions we get \bea C_{1}\cdot
C_{2}\geq n_{1}n_{2}\{1-\frac{1}{2}(f_{\lambda^{a}}(n_{1})
+f_{\lambda^{b}}(n_{2}))+\frac{1}{2}(\frac{\nu}{n_{2}}-\frac{\mu}{n_{1}})^{2}\}
\eea Since $f_{\lambda^{a}}(n)=\min_{\lambda\in
\Gamma_{8}}(\lambda-\frac{\lambda^{a}}{n})$ therefore
$f_{\lambda^{a}}(n)\leq (\frac{\lambda^{a}}{n})^{2}$. The dominant
weights satisfy the condition \bea
\sum_{i=1}^{8}a_{i}\lambda_{i}=2(\lambda_{1}+\lambda_{7})+3(\lambda_{2}+\lambda_{8})+4(\lambda_{3}+\lambda_{6})+5\lambda_{4}+6\lambda_{5}\leq
n\,. \eea if we define the matrix $B$ as follows \bea
B_{ij}&=&2a_{i}a_{j}\,,i\neq j\,,\\ \nn &=&a_{i}^{2}\,,i=j\,. \eea
Then we see that $B_{ij}\lambda_{i}\lambda_{j}\leq n^{2}$ and
since $B_{ij}-(A^{-1}_{E_{8}})_{ij}\geq 0$ therefore it follows
that $\lambda^{2}\leq n^{2}$ if $\lambda$ is a level $n$ dominant
weight. Thus we get $f_{n}(\mu)\leq 1$. Which therefore implies that,
\bea C_{1}\cdot C_{2}&\geq&
n_{1}n_{2}\{1-\frac{1}{2}(f_{\lambda^{a}}(n_{1})
+f_{\lambda^{b}}(n_{2}))+\frac{1}{2}(\frac{\nu}{n_{2}}-\frac{\mu}{n_{1}})^{2}\}\,,\\
\nn &\geq& n_{1}n_{2}\{1-\frac{1}{2}(f_{\lambda^{a}}(n_{1})
+f_{\lambda^{b}}(n_{2}))\}\geq 0. \eea From the fact that
$(\lambda^{a})^{2}\leq n^{2}$ it follows that $\lambda^{a}$ is the
vector of least length in its class and therefore \bea
f_{n}(\lambda)=\frac{(\lambda^{a})^{2}}{n^{2}}\,\,,\,\forall\,\,\lambda\in
\lambda^{a}+n\Gamma_{E_{8}}. \eea
Then from Eq(\ref{genuseq}) it follows that
\bea
g_{c}(n\lambda^{a})=1+\mbox{$\frac{n(n-1)}{2}$}-\mbox{$\frac{(\lambda^{a})^{2}}{2}$}\,.
\eea

\section*{Appendix B}
The Calabi-Yau space $X$ whose compact divisor is the rational elliptic surface,${\cal B}_{9}$,
is the total space of the canonical  bundle over ${\cal B}_{9}$.
The mirror manifold $Y$ is given by \cite{estring,SZ,HI}\bea
y^{2}&=&x^{3}+f_{4}(z)x+g_{6}(z)\,,\\ \nn uv&=&z-z_{*}\,.  \eea
Where $z_{*}$ is a function of the K\"ahler moduli of $X$.
The first equation defines an elliptic fibration with twelve
singular fibers. Apart from the fact that the base space given by
$z$ is non-compact (and therefore the total manifold is hyper
K\"ahler) the geometry of the singular fibers is exactly the same
as the compact $\frac{1}{2}$K3. The twelve degenerate fibers can
be chosen to have the following charges\footnote{With respect to a
particular choice of paths \cite{7-brane,weyl}.}, \bea
[1,0][1,0]\cdots [1,0][2,-1],[-1,-1],[-1,2]\,.  \eea It was shown
in \cite{7-brane,weyl} that this configuration is invariant under
$\sl2z$ transformations. The effect of this $\sl2z$ transformation
on the charges was determined in \cite{HI} and is given in
Eq(\ref{T}).

The transformations $G_{\pm}$ are more interesting and correspond
to generalized Weyl transformation. Consider the lattice $H_{*}(X,
\bbbz)=H_{0}(X,\bbbz)\oplus H_{2}(X,\bbbz)\oplus H_{4}(X,\bbbz)$.
A codimension one lattice in $H_{2}(X)$, orthogonal to the
canonical class, is isomorphic to the root lattice of the affine
$E_{8}$ algebra. One way of seeing this is that we have the
ordinary $E_{8}$ algebra which is enhanced to affine $E_{8}$ by
the canonical class , $F$, since $F^{2}=0$. Another way of obtaining the affine $E_{8}$
is to find another element of $H_{*}(X)$ which squares to zero. It
is easy to see that only other such element is the generator of
$H_{0}(X)$. Thus we have two different copies of affine $E_{8}$
root lattice in $H_{*}(X)$. Thus the most general root of
$H_{*}(X)$ is given by (using the same notation as in section 3)\bea
\widehat{\alpha}_{n,m}=(0,\{\alpha,0,n\},m)\,,\,\,\alpha^{2}=2\,,\,\,n,m\in
\bbbz\,.  \eea It is clear that $\langle \widehat{\alpha}_{n,m},\widehat{\alpha}_{n,m}\rangle =-2$
and therefore $\langle\widehat{V},\widehat{V}\rangle=\langle V,V\rangle$ where
$\widehat{V}=V+\langle V,\widehat{\alpha}_{n,m}\rangle\,\widehat{\alpha}_{n,m}$.
Two successive transformation using first the root
$\widehat{\alpha}_{n,m}$ and then $\widehat{\alpha}_{n,0}$
gives\bea
V:=(0,\{\lambda,d,s\},k)\mapsto \widehat{V}=(0,\{\lambda,d,s\},k+m(-\lambda\cdot\alpha+dn)\}\,.\eea
Thus we see that if $\lambda \in h\Gamma_{E_{8}}$ then $k$ can be changes in units
of $\mbox{gcd}(h,d)$ by generalized Weyl transformations.

\section*{Appendix C}
\bea
\theta_{2}(\tau)&:=&\sum_{n\in \bbbz}\,q^{\frac{(n-\frac{1}{2})^{2}}{2}}\,,\,\,\,
\theta_{3}(\tau):=\sum_{n\in \bbbz}\,q^{\frac{n^{2}}{2}}\,,\,\,\,
\theta_{4}(\tau):=\sum_{n\in \bbbz}\,(-1)^{n}q^{\frac{n^{2}}{2}}\,,\\ \nn
\theta_{2}^{4}(\mbox{$-\frac{1}{\tau}$})&=&-\tau^{2}\theta_{4}^{4}(\tau)\,,\,\,\,
\theta_{3}^{4}(\mbox{$-\frac{1}{\tau}$})=-\tau^{2}\theta_{3}^{4}(\tau)\,,\,\,\,
\theta_{4}^{4}(\mbox{$-\frac{1}{\tau}$})=-\tau^{2}\theta_{2}^{4}(\tau)\,.
\eea
\bea
E_{2}(\tau)&=&1-24\sum_{k=1}^{\infty}\sigma_{1}(k)q^{k}\,,\,\,\,\\ \nn
E_{4}(\tau)&=&1+240\sum_{k=1}^{\infty}\sigma_{3}(k)q^{k}\,,\,\,\,
E_{4}(\mbox{$-\frac{1}{\tau}$})=\tau^{4}E_{4}(\tau)\,,\,\,\sigma_{s}(k):=\sum_{n|k}n^{s}\,.
\eea
\bea
P_{2,0}(\tau)&=&E_{4}(2\tau)\,,\,\,\,
P_{2,\lambda^{1}}(\tau)=\frac{E_{4}(\frac{\tau}{2})+E_{4}(\frac{\tau}{2}+\frac{1}{2})}{2}-E_{4}(2\tau)\,,\\\nn
P_{2,\lambda^{2}}(\tau)&=&\frac{E_{4}(\frac{\tau}{2})-E_{4}(\frac{\tau}{2}+\frac{1}{2})}{2}\,,\,\,\,\lambda^{1}\cdot\lambda^{1}=4\,,\,\,\lambda^{2}\cdot \lambda^{2}=2\,.
\eea

\parskip=0pt plus 2pt

\end{document}